\documentclass[aps,superscriptaddress,showpacs,twocolumn]{revtex4}
\usepackage[dvips]{graphics}

\begin{document}
\title{Relativistic Weierstrass random walks}

\def \beq {\begin{equation}}
\def \eeq {\end{equation}}

\author{Alberto Saa}\email{asaa@ime.unicamp.br}
\affiliation{Departamento de Matem\'atica Aplicada, UNICAMP,  13083-859 Campinas, SP, Brazil}
\author{Roberto Venegeroles}\email{roberto.venegeroles@ufabc.edu.br}
\affiliation{Centro de Matem\'atica, Computa\c c\~ao e Cogni\c c\~ao, UFABC, 09210-170 Santo Andr\'e, SP, Brazil}

\date{\today}

\begin{abstract}
The Weierstrass random walk is a paradigmatic Markov chain giving rise to a L\'evy-type superdiffusive  behavior. It is well known that  Special
Relativity prevents the arbitrarily high velocities necessary to
establish a superdiffusive behavior in any process occurring in Minkowski spacetime, implying, in particular, that
any relativistic Markov chain describing spacetime phenomena must be essentially Gaussian.
Here, we introduce a simple  relativistic extension of the Weierstrass
random walk and
  show  that  there must exist   a transition time $t_c$
delimiting  two qualitative distinct dynamical regimes: the
(non-relativistic) superdiffusive
 L\'evy flights, for $  t < t_c$, and the usual (relativistic) Gaussian
diffusion, for $t>t_c$.
Implications of this  crossover
between different
diffusion regimes
   are discussed for some explicit
examples.
The study of such an explicit and simple Markov chain can shed some light on
 several  results obtained in   much more
involved contexts.
\end{abstract}


\maketitle

\section{Introduction}

Relativistic Brownian motion is an interesting and active area of research nowadays\cite{Review}. The dynamical behavior of any relativistic
process is, of course, strongly constrained  by the main
 properties of the relativistic kinematics, namely
the speed of light $c$ as the maximum possible physical
velocity, the invariance under
Lorentz transformations, and
the causal structure associate with the light-cone. Some well established
non-relativistic dynamical behaviors are simply incompatible with the
principle of Special
Relativity. This is specifically the case of the so-called L\'evy flights \cite{Levy}, where very rare events with arbitrarily high velocities give rise to a superdiffusive regime characterized by a power law
\beq
\label{sqr}
\langle X^2(t)\rangle \propto t^\mu,
\eeq
where $\mu >1$ is the corresponding anomalous diffusion exponent.
However, Special Relativity does  not allow such arbitrarily high velocities, preventing the appearance of such superdiffusive regime in Markov chains involving
spacetime events.
These points are studied, for instance,  in \cite{JETP} by using
  relativistic versions of the Fokker-Planck equation  and   of the fluctuation-dissipation theorem, leading to
a fractional-derivative
extension of the diffusion equation. An earlier analysis of  relativistic
random walks can be found in \cite{Wall}. In \cite{PRD}, a generalized Wiener process avoiding
superluminal propagation is introduced, giving rise to a  non-Markovian
 relativistic diffusion process. The influence of the spacetime causal structure  on dynamical processes, namely the implication of the presence of an event horizon, was investigated
recently  in \cite{MotterSaa}.

We consider here a simple relativistic extension of the Weierstrass random walk in order to shed some light on several relativistic aspects of  L\'evy flights.
We remind that the usual
one-dimensional Weierstrass random walk\cite{Weier} corresponds to a Markov chain  governed by the following
probability density function
\beq
\label{PDF}
\psi(x) = \frac{a-1}{2a}\sum_{n=0}^\infty a^{-n}\left(
\delta\left(x+vJ_n\right) + \delta\left(x-vJ_n\right)
 \right),
\eeq
with the ``jump'' function
\beq
\label{weier}
J_n = J_n^{\rm (NR)} = b^n,
\eeq
where $a>1$ and $b>1$ are dimensionless constants and $v>0$ gives the scale
of the jumps. A particle moving according to (\ref{PDF}) and (\ref{weier})
can perform
jumps, in both directions, with magnitude $v, bv , b^2v, b^3v, \dots$, and  with
probability, respectively, given by $(a-1)/a, (a-1)/a^2, (a-1)/a^3, (a-1)/a^4, \dots$.
We consider here that each step $\ell$ lasts for a given and fixed time interval and, hence,
the asymptotic dynamics for large times and for large number of steps
 are identical.
One can think the constant $v$ as the velocity acquired by the particle,
just prior to the jump,
by some unspecific microscopic mechanism.
Notice that one can indeed define a continuous-time version of the Weierstrass
random walk\cite{CT}, but for our purposes here, the simple
Markov chain  governed (\ref{PDF}) is
enough.

We are mainly interested in the anomalous diffusion process associated to the
Weierstrass
random walk. A brief review of the main results concerning this
topic is necessary here.
We wish to characterize the large $L$ behavior
of $\left\langle X^2(L) \right\rangle$,
where
\beq
X(L) = x_1+x_2+x_3+\cdots +x_L,
\eeq
with $x_\ell$ being the size of the $\ell^{\rm th}$ jump, with probability density
function  given by (\ref{PDF}) and (\ref{weier}). Since
$x_\ell$ are independent random variables for different $\ell$, we will have
\beq
\left\langle X^2(L) \right\rangle = L \langle x^2\rangle.
\label{diff}
\eeq
Notice that
\beq
\label{quadr}
\langle x^2\rangle = \int_{-\infty}^\infty x^2\psi(x)\, dx = \frac{a-1}{a} v^2
\sum_{n=0}^\infty \left(\frac{b^2}{a}\right)^n,
\eeq
from where we
  have that, for $b^2 < a$, $\langle x^2\rangle$ is finite, equation (\ref{diff})
reduces to the usual diffusion process with $\mu=1$, and, thanks to the central limit theorem, the total probability density function describing the
passage from $\ell=1$ to $\ell=L$ will be very close
to a Gaussian.
For $b^2 \ge a$, on the other hand, $\langle x^2\rangle$ diverges,
the central limit theorem cannot by applied anymore and the associated distributions
will not be Gaussian,
leading to an anomalous  diffusion exponent $\mu>1$.
We can  use the heuristic approach of \cite{Kutner} in order to evaluate $\mu$ for this case as well, which correspond
to $\alpha \le 2$, where
\beq
\alpha = \frac{\log a}{\log b}.
\eeq

The key idea is that, according to (\ref{PDF}) and (\ref{weier}), if one considers a large number $L$ of steps, the most frequent steps will have magnitude $v$. Steps with magnitude
$bv$ will be $1/a$ less frequent than the previous one. In general,
steps with magnitude $b^{n+1}v$ will be $1/a$ less frequent than those ones
with magnitude $b^nv$. Also, if we consider large, but finite,
 number   of steps,
the summation (\ref{PDF}) will be effectively truncated at a
 given value $n=n_{\rm max}$. Let us
consider the following succession of steps, which exhibits the required
hierarchy of jumps,
\beq
L = a^{n_{\rm max}} + a^{n_{\rm max}-1} + a^{n_{\rm max}-2} + \cdots + 1,
\eeq
corresponding to $a^{n_{\rm max}}$ steps with magnitude $v$, $a^{n_{\rm max}-1}$
with with magnitude $bv$, and so on. For large $n_{\rm max}$, we have
\beq
\label{L}
L = \sum_{j=0}^{n_{\rm max}} a^{n_{\rm max}-j} \approx \frac{a}{a-1} a^{n_{\rm max}}.
\eeq
Yet for large but
  finite $n_{\rm max}$, we can estimate $\langle x^2\rangle$
as
\beq
\label{est}
\frac{\langle x^2\rangle}{v^2} \approx \frac{a-1}{a}
\sum_{j=0}^{n_{\rm max}} \left(\frac{b^2}{a}\right)^j.
\eeq
For $\alpha > 2$, as we already know, we have the Gaussian result since
\beq
\frac{\langle x^2\rangle}{v^2} \approx \frac{a-1}{a-b^2} < \infty,
\eeq
which coincides with the exacted  result evaluated from (\ref{quadr}).
For $\alpha < 2$, on the other hand, equation (\ref{est}) implies that
\beq
\label{quad1}
\frac{\langle x^2\rangle}{v^2} \approx
\left(\frac{a-1}{a} \right)^\frac{2}{\alpha}
\frac{a}{b^2-a}L^{\frac{2}{\alpha}-1},
\eeq
where (\ref{L}) was used,
leading to a superdiffusive $(\mu = 2/\alpha >1)$ behavior characterized by
\beq
\label{quad11}
\frac{\left\langle X^2(L) \right\rangle}{v^2}\approx
\left(\frac{a-1}{a} \right)^\frac{2}{\alpha}
\frac{a}{b^2-a}L^{\frac{2}{\alpha}}.
\eeq
The relativistic kinematics, however, will change  dramatically this scenario.

\section{The relativistic walk}

The microscopical origin of the jumps in a non-relativistic random walk
is not relevant from the dynamical point of view. Provided that the Markov
property holds, {\em i.e.}, the position $X(\ell)$ of the
 system at a given step $\ell$ depends only on
 the position at the previous
 step $X(\ell-1)$,
and that successive jumps are independent random
variables,   the standard approach to lead with random walks \cite{Review}
can be applied. Let us, nevertheless,
suppose that the jumps are due to microscopical collisions as in the
historical example of Brownian motion.
The hierarchy of jumps  $v, bv , b^2v, b^3v, \dots$ is, in
  the nonrelativistic case,
associated with a similar hierarchy of momentum transfers in the collisions
$p, bp , b^2p, b^3p, \dots$,
with $p=m_0v$, where $m_0$ is the particle rest mass.
However, according to Special Relativity,   the velocity and
the momentum of a particle should obey
\beq
p=\frac{m_0v}{\sqrt{1-\beta^2}},
\eeq
where $\beta = v/c$, with profound implications for the jump hierarchy.
  Assuming that the same hierarchy of momentum transfers is present in the relativistic case, we will have the following jump function
\beq
\label{relatweier}
J_n = J_n^{\rm (R)} = \frac{b^n}{\sqrt{1+ \beta^2 b^{2n}}},
\eeq
instead of (\ref{weier}).
The hierarchy of relativistic jumps will be
 $\{J_nv\}$, occurring in the dynamics,
respectively,
 with probability $\{(a-1)/a^{n+1}\}$, with $n=0,1,2,\dots$.
Notice that, in frank contrast with the nonrelativistic case, for large $n$, one has $J_nv \approx c$, meaning that
there will be no arbitrarily large jumps in the relativistic case, in agreement
with the fact that  no acquired velocity by any microscopical
mechanism  can exceed $c$.
 The first conclusion we can draw from this relativistic extension of the
 Weierstrass
random walk   is that the whole process
must be Gaussian, since, in this case, we have
\beq
\label{quadrel}
\frac{\langle x^2\rangle}{v^2} = \frac{a-1}{a}
\sum_{n=0}^\infty \frac{(b^{2}/a)^n}{ 1+\beta^2b^{2n} } < \infty,
\eeq
for any $\beta>0$, implying the usual diffusion with $\mu=1$, irrespective
of the value of $\alpha$.

\begin{figure}[t]
\resizebox{1\linewidth}{!}{\rotatebox{0}{\includegraphics*{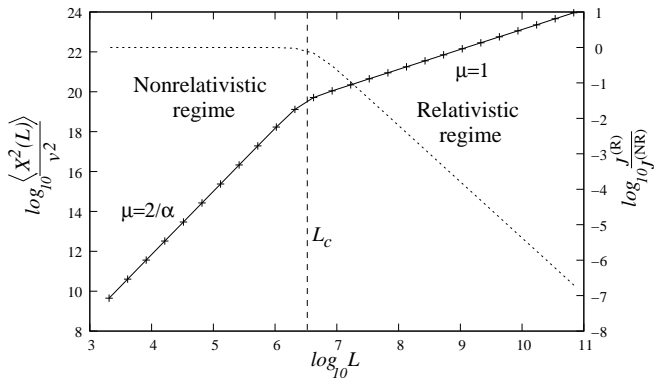}}}
\caption{Diffusion in the relativistic Weierstrass random walk. The solid
line (left scale) corresponds to (\ref{quadrel2}), which, for large $n_{\rm max}$, is well approximated by
 (\ref{eqtot}). The dotted line (right scale) is the ratio between the non-relativistic (\ref{weier}) and relativistic (\ref{relatweier}) jump functions. It is clear the appearance of a  crossover between the different diffusion regimes
 near $L_c$ given by (\ref{Lc}).
 For $L<L_c$, we have L\'evy-type superdiffusive  behavior, while for
 $L>L_c$ the dynamics settle
into a Gaussian diffusion.
  This plot corresponds to the case where
 $a=2$, $b=3$, and $\beta = 10^{-10}$.}
\label{fig1}
\end{figure}

Typical non-relativistic situations are characterized by a small $\beta$.
For such cases,  from
(\ref{quadrel}), one realizes that there should exist a critical
 value $n_c$ such
that, for $n \ll n_c$, the summand of (\ref{quadrel}) is essentially the same one of
the nonrelativistic case. If it is possible to choose a large
$n_{\rm max}\ll n_c$, one could
 apply the same heuristic approach of last section,
implying in a superdiffusive behavior with $\mu = 2/\alpha > 1$.
(We assume hereafter that $0<\alpha < 2$.)
This occurs because $L$ is large enough to justify the averages of last section, but it is still small enough to guarantee that the large $n$ events that would
imply the relativistic regime are extremely rare and will not contribute
effectively  to the
averages. In other words, the system needs some time to realize that it is indeed relativistic! If we allow for $n_{\rm max}\gg n_c$, we will have
in (\ref{quadrel})
 the convergent
relativistic summation, implying the usual diffusion with exponent
$\mu=1$. It is clear that
\beq
n_c = -\frac{\log\beta}{\log b} =\alpha \frac{\log \beta^{-1}}{\log a},
\eeq
leading to
\beq
\label{Lc}
L_c = \frac{a}{a-1}\frac{1}{\beta^\alpha},
\eeq
where (\ref{L}) was used.
 If $\beta$ is small, as one expects
  in the typical nonrelativistic problems, $L_c$ will
be large. For  $L\ll L_c$, the system behaves as in the nonrelativistic
regime and exhibits the properties of a L\'evy flight with
$\mu=2/\alpha$. On the other hand, for
 $L\gg L_c$, the system change its behavior to the relativistic regime,
 characterized by ordinary diffusion with $\mu=1$.
 Such crossover
between different
diffusion regimes,
 depicted in Fig. \ref{fig1}, is compatible
with the results of \cite{Crossover}. In fact, our model can be viewed
 as a simple microscopical realization for the generalized Fokker-Planck equation
considered there.

 We can apply the same heuristic approach of last section to the
 relativistic case. In particular, we have for a finite and large
 $n_{\rm max}$
 \beq
 \label{quadrel2}
 \frac{\left\langle X^2(L) \right\rangle}{v^2}
  \approx \frac{a-1}{a} L
\sum_{n=0}^{n_{\rm max}} \frac{(b^{2}/a)^n}{ 1+\beta^2b^{2n} },
 \eeq
 leading to the following generalization of (\ref{quad11}) for $\beta\ne 0$
(see the Appendix for details)
\begin{widetext}
\beq
\label{eqtot}
\frac{\left\langle X^2(L) \right\rangle}{v^2}  \approx  \frac{1}{\log (b^2/a)}
\left[
\left(\frac{a-1}{a}\right)^\frac{2}{\alpha}
L^\frac{2}{\alpha}
  F\left(1,1-\frac{\alpha}{2}; 2-\frac{\alpha}{2};
-\left(\frac{L}{L_c} \right)^\frac{2}{\alpha} \right)
- \frac{a-1}{a} LF\left(1,1-\frac{\alpha}{2}; 2-\frac{\alpha}{2};
-\beta^2  \right)
\right]
,
\eeq
\end{widetext}
where $F(a,b;c;z)$ stands for the standard  hypergeometric function\cite{Abramowitz}.
Using that $F(a,b;c; 0) = 1$, we have
from (\ref{eqtot})  an anomalous
diffusion process with $\mu=2/\alpha$.
for large $L$ obeying
 $L\ll L_c$.
  On the other hand, for $L\gg L_c$,   we have (see the Appendix for details)
\beq
\label{asym}
F\left(1,1-\frac{\alpha}{2}; 2-\frac{\alpha}{2};
-\left(\frac{L}{L_c} \right)^\frac{2}{\alpha} \right) \propto \left(\frac{L}{L_c} \right)^{1-\frac{2}{\alpha}},
\eeq
leading to the usual Gaussian diffusion
\beq
\frac{\left\langle X^2(L) \right\rangle}{v^2}\propto L,
\eeq
for $L\gg L_c$.

 \section{Discussion}

The results of the preceding sections can be summarized as follows.
Suppose we have a Weierstrass random walk model with typical
 velocity $v$, implying in L\'evy flights
characterized by an anomalous diffusion exponent $\mu > 1$. Then,
relativistic effects imply that, after a certain critical number
of steps $L_c\approx (c^2/v^2)^\frac{1}{\mu}$, the system loses it
anomalous diffusion properties and the dynamics necessarily settle
into a Gaussian diffusion. In order to estimate the order of magnitude
of these relativistic effects, let us
  associate   the kinetic energy of the walking particle,
and thus $v$, with the typical  thermal energy $k_BT$.
For non relativistic situations, where $k_BT$ is small if compared with
$m_0c^2$, we will have     $m_0v^2/2 \approx k_BT$, leading to
\beq
L_c \approx \left(\frac{m_0c^2}{2k_BT}\right)^{\frac{1}{\mu}}.
\eeq

As our first explicit example, let us consider a system composed by
helium atoms, for which  $m_0c^2/k_B \approx 4.36\times 10^{13}$K.
For such a system at room temperature
($T\approx 300 $K), a ballistic $(\mu=2)$ L\'evy flight originated in
a Weierstrass random walk, will become  Gaussian due to
relativistic effects after $L_c \approx 2.7\times 10^{5}$ steps.
 Helium
atoms at the surface of the sun ($T\approx 5\times 10^{3}$K) can
experiment ballistic L\'evy flights in a Weierstrass random walk for no more than
$L_c\approx 6.6\times 10^4$ steps. In the interior of the sun ($T\approx 5\times 10^{6}$K),   only $L_c\approx 2,000$ steps will be enough for the
the dynamics settle into a  essentially
Gaussian regime.

Heavier particles or bodies will, naturally, lead to larger values for $L_c$.
Let us take, for instance, the case of an {\em Escherichia coli} bacterium, for which $m_0=665$ femtograms \cite{Ecoli}, leading to
$m_0c^2/k_B \approx 4.33\times 10^{24}$K. At room temperature, $L_c\approx
4.1\times 10^{15}$ for an {\em Escherichia coli}
undergoing a L\'evy flight originated in
a Weierstrass random walk with diffusion exponent close to those
ones observed experimentally\cite{Micelle} in systems of breakable micelles
 $(\mu\approx 1.4)$. One realizes that such
microscopic bodies can indeed experience much longer
anomalous diffusion process than atomic scale particles.

\acknowledgments

This work was supported by the Brazilian agencies CNPq and FAPESP.

\appendix
\section{}

We can approximate (\ref{quadrel2}) by an integral by using
\beq
\label{integr}
\sum_{n=0}^{n_{\rm max}} \frac{(b^{2}/a)^n}{ 1+\beta^2b^{2n} }
  \approx  \frac{1}{2\log b}
\int_{1}^{b^{2n_{\rm max}}} \frac{ w^{-\frac{\alpha}{2} } }{ 1+\beta^2 w} \,dw ,
\eeq
where  variable $w = b^{2n}$ was introduced and
  $\alpha$ is, in general,  an irrational.
For $|\beta^2 w|<1$, we can introduce the following series expansion
\beq
\label{expans}
\frac{1}{1+\beta^2 w} = \sum_{n=0}^\infty (-\beta^2  w)^n,
\eeq
and the integral (\ref{integr}) will  be written as
\begin{eqnarray}
\label{hyper}
&&
\int  \frac{ w^{-\frac{\alpha}{2} } }{ 1+\beta^2 w} \,dw  =
w^{1-\frac{\alpha}{2}}\sum_{n=0}^\infty \frac{(-\beta^2 w)^n}{n+1-\frac{\alpha}{2}}  \\
&& =  \frac{2}{2-\alpha} w^{1-\frac{\alpha}{2}} F\left(1,1-\frac{\alpha}{2} ; 2-\frac{\alpha}{2} ;
-\beta^2 w \right), \nonumber
\end{eqnarray}
where $F(a,b;c;z)$ is the
standard hypergeometric function \cite{Abramowitz}. Notice that
the hypergeometric functions have a single valued analytical extension over the
entire complex plane, with the only exception of the positive real axis for
$z\ge 1$ \cite{Abramowitz}, justifying the use of (\ref{hyper}) for the evaluation of the
integral (\ref{integr}), which limits, in fact, do not belong to the
region where the expansion (\ref{expans}) converges.
The integral    (\ref{integr}) may also be evaluated by exploring
the hypergeometric function identities\cite{Abramowitz}
\beq
\frac{d}{dz}\left(z^{c-1}F(1,b;c;z) \right) = (c-1)z^{c-2}F(1,b;c-1;z),
\eeq
and $ F(1,b;b;z) = (1-z)^{-1}$,
valid for any $b$ and $c$.
Equation (\ref{eqtot}) follows straightforwardly from (\ref{hyper}).

The evaluation of   (\ref{eqtot})  for $L\gg L_c$
 requires an asymptotic analysis for the
hypergeometric function. By using, for instance, the identity 15.3.8 of
\cite{Abramowitz},  we have
\begin{widetext}
\beq
F\left(1,1-\frac{\alpha}{2} ;2-\frac{\alpha}{2} ;1-z\right) = \frac{2-\alpha}{\alpha}z^{-1}
F\left( 1,1 ;\frac{\alpha}{2}-1  ; z^{-1}\right) +
\frac{\pi \left( 1-\frac{\alpha}{2}\right)}{\sin\pi\frac{\alpha}{2}}z^{\frac{\alpha}{2}-1}
F\left( 1-\frac{\alpha}{2},1-\frac{\alpha}{2} ;1-\frac{\alpha}{2}  ; z^{-1}\right),
\eeq
\end{widetext}
implying that, for large $z$ and $0<\alpha <2$,
\beq
F\left(1,1-\frac{\alpha}{2}; 2-\frac{\alpha}{2};
-z \right) \approx
\frac{\pi \left( 1-\frac{\alpha}{2}\right)}{\sin\pi \frac{\alpha}{2}}z^{\frac{\alpha}{2}-1},
\eeq
leading finally to (\ref{asym}).

\end{document}